\documentclass[twoside,reqno]{lt7-proc}
\usepackage{epsfig,cite}
\usepackage{amssymb,amsmath}
\usepackage{times}
\usepackage{graphics}
\usepackage{amsfonts}
\usepackage{shadow}
\usepackage{float}
\begin{document}
\sloppy \raggedbottom
\setcounter{page}{1}
%%%%%%%%%%%%%%%%%%%%%%%%%%
%%%%%%%%%%%%%%%%%%%%%%%%%%%%%%%%%%%%%%%%%%%%%%%%%%%%%%%%%%%%%%%%%
\newcommand{\beq}{\begin{equation}}
\newcommand{\eeq}{\end{equation}}
\newcommand{\beqa}{\begin{eqnarray}}
\newcommand{\eeqa}{\end{eqnarray}}
\newcommand{\nn}{\nonumber \\}
\newcommand {\np}[1]{{\mbox{\textrm{:}\,}{#1}{\,\textrm{:}}} }
\def \e {\mathrm{e}}
\def \edge {\mathrm{edge}}
\def \eps {\varepsilon}
\def \k {\kappa}
\def \l {\lambda}
\def \la {\langle}
\def \ra {\rangle}
\def \s {\sigma}
\def \t {\tau}
\def \B {{\mathcal B}}
\def \C {{\mathbb C}}
\def \R {{\mathbb R}}
\def \Z {{\mathbb Z}}
\def \ch {\mathrm{ch}}
\def \el {\mathrm{el}}
\def \qh {\mathrm{qh}}
\def \qp {\mathrm{qp}}
\def \z {\zeta}
\def \L {\underline{\Lambda}}
\def \D {\Delta}
\def \I {{\mathbb I}}
\def \Pf {\mathrm{Pf}}
\def \Im {\mathrm{Im} \, }
\def \Re {\mathrm{Re} \, }
\def \sh {\mathrm{sh}}
\def \tx {\tilde{x}}
\def \mod {\ \mathrm{mod} \ }
\def \H {{\mathcal H}}
\def \uu {{\widehat{u(1)}}}
%%%%%%%%%%%%%%%%%%%%%%%%%%%%%%%%%%%%%%%%%%%%%%%%%%%%%%%%%%%%%%%%%%
%\bibliographystyle{utphys}
\newpage
\setcounter{figure}{0}
\setcounter{equation}{0}
\setcounter{footnote}{0}
\setcounter{table}{0}
\setcounter{section}{0}

% Title, authors and addresses

% use the thanks command within \title, \author or \address for
%footnotes: % \title{Title} or  \title{Title\thanks{...}}

\title{Topological Quantum Computation with the universal $R$ matrix for Ising anyons}

\runningheads{L.S. Georgiev}{{Topological Quantum Computation} }

\begin{start}

% \author{Name1}{aff.label1},
%\coauthor{Name2}{aff.label2},
%\coauthor{Name3}{aff.label3}
%\address{Address1}{aff.label1}
%\address{Address2}{aff.label2}
%\address{Address3}{aff.label3}

\author{Lachezar S. Georgiev}{1,2},

\address{Institute for Nuclear Research and Nuclear Energy \\
Bulgarian Academy of Sciences \\
72 Tzarigradsko Chaussee, 
1784 Sofia, Bulgaria}{1}

\address{Institut f\"ur Mathematische Physik, Technische Universit\"at Braunschweig,
Mendelssohnstr. 3, 38106 Braunschweig, Germany}{2}

%you may repeat \coauthor as many times as you need
%names may have more than one aff.label, e.g.,
%\coauthor{Name2}{aff.label2,aff.label3},

\begin{Abstract}
We show that the braid-group extension of the monodromy-based 
topological quantum computation scheme of Das Sarma et al. can be understood 
in terms of the universal $R$ matrix for the Ising model giving similar results 
to those obtained by direct analytic continuation of multi-anyon Pfaffian wave functions.
It is necessary, however, to take into account the projection on spinor states with 
definite total parity  which is responsible for the topological entanglement in
the Pfaffian topological quantum computer.
\end{Abstract}
\end{start}

%%%%%%%%%%%%%%%%%%%%%%%%%%%%%%%%%%%%%%%%%%%%%%%%%%%%%%%%%%%%%
% The main text of your paper                               %
%%%%%%%%%%%%%%%%%%%%%%%%%%%%%%%%%%%%%%%%%%%%%%%%%%%%%%%%%%%%%
\section{Introduction}
Quantum Computation \cite{nielsen-chuang} in general is expected 
to allow us to solve
computational problems that are hard to attack by classical methods.
Unfortunately, this exponential speed-up over classical computation
has not been demonstrated  so far because of the overwhelming noise and
decoherence due to the coupling between the qubits and their
environment. The Topological Quantum Computation (TQC) 
\cite{kitaev-TQC,preskill-TQC,sarma-freedman-nayak} is an intriguing 
proposal to use
the braiding operations of non-Abelian quasiparticles in certain
strongly correlated electron systems, such as the fractional quantum Hall (FQH)
liquids, as quantum gates.
The main advantage of using this highly complicated approach is that
the encoding of quantum information, as well as the representation of
the corresponding quantum operations, are naturally immune against noise
and decoherence due to the presence of a bulk energy gap in the
excitations spectrum. This has been called topological protection
of quantum information and quantum operations because quantum
information is naturally encoded in topological quantum numbers
and then quantum gates must be implemented by topologically nontrivial 
operations.

One of the remarkable theoretical achievement in this direction was the
TQC scheme of Das Sarma et al. \cite{sarma-freedman-nayak}
in which the qubit has been constructed by 4 non-Abelian Ising anyons
localized on 4 antidots in a Pfaffian FQH liquid and the qubit measurement
was implemented electronic by Mach--Zehnder interferometry.
In this scheme the computational basis for quantum computation was realized in
terms of 4-point correlations functions of the critical two-dimensional Ising model
and a logical NOT gate has been implemented by a monodromy transformation.

The monodromy-based TQC scheme of Ref.~\cite{sarma-freedman-nayak} has been extended \cite{TQC-PRL,TQC-NPB}to include braid-group operations  by using the exchange 
matrices for 4-quasiholes  Ising-model wave functions in the Pfaffian FQH state.
Unfortunately it turns out that the representation of the (infinite)
braid group $\B_4$ is actually finite and as a result not all single-qubit gates
could be realized simply by braiding. 
In addition to the single-qubit construction, an important two-qubit generalization
has been proposed in Refs.~\cite{TQC-PRL,TQC-NPB}, in terms of 6 Ising anyons,
and two-qubit gates have been implemented by braiding operations from the Ising-model representation of the braid group $\B_6$. The elementary braid matrices for 6 quasiholes
have been computed explicitly from the 6-quasiholes Pfaffian wave functions
and a completely topologically protected construction of the 
Controlled-$Z$ and Controlled-NOT gates has been given \cite{TQC-PRL,TQC-NPB}.

In this Report we will demonstrate that the above braid matrices could be generated
from the universal $R$ matrix for the Ising model, obtained before 
\cite{nayak-wilczek,ivanov}, by introducing 
an appropriate projector taking into account the fermion parity conservation in the Ising-model correlation functions. This projection turns out to be the origin of the topological entanglement
which is characteristic for the Pfaffian topological quantum computer.
%%%%%%%%%%%%%%%%%%%%%%%%%%%%%%%%%%%%%%%%%%%%%%%%%%%%
\section{Quantum computation in general}
\label{sec:general}
%%%%%%%%%%%%%%%%%%%%%%%%%%%%%%%%%%%%%%%%%%%%%%%%%%%%%
The quantum bit of information, called the qubit, is a quantum state
which belongs to a two-dimensional complex vector space usually represented in an 
orthonormal basis $\{|0\ra, \ |1\ra\}$ as a normalized coherent superposition
\[
 |\psi\ra = \alpha|0\ra +\beta |1\ra, \quad \mathrm{where} \quad \alpha, \ \beta \in \C 
\quad \mathrm{and} \quad  |\alpha|^2+|\beta|^2=1.
\]
In any concrete quantum computation approach we have to first define the computational basis
$|0\ra$,  $|1\ra$  and then initialize the system by constructing e.g., $|\psi\ra= |0\ra$ or  $|\psi\ra= |1\ra$ (or some known superposition state).
The measurement is understood as a standard orthogonal projectors family defining the projected state after measurement and the corresponding probability \cite{nielsen-chuang}.
The multiple qubits are constructed as tensor products of the individual qubits so that, e.g.,
 the $n$-qubit space of states is (the projective space of)
\[
\H^n = \otimes^n\C^2  \simeq \C^{2^n}. 
\]
The initialization of the multi-qubit system in quantum computation is usually done by 
preparing the system in a specific state in the multi-qubit Hilbert space
\[
010011000...01 \quad \mathop{\to}\limits^{\mathrm{QC}} \quad |010011000...01\ra \in \H^n .
\]
Any quantum computation could be performed by applying to the input multi-qubit state a 
finite sequence of quantum gates,
which are represented by unitary operators over $\H^n$ and
the output of the computation which is obtained after measurement is classical data
that could be written on paper.
The exponential speed-up of quantum computation is due to the quantum entanglement
and to a phenomenon called quantum parallelism \cite{nielsen-chuang}.

Any quantum gate can be approximated \cite{nielsen-chuang}, 
with arbitrary precision with respect to 
$
E(U,V)\equiv \max_{|\psi\ra} || \left( U-V \right)|\psi\ra||,
$
by products of 3 universal gates: $H$, CNOT and $T$. These gates are called discrete set of universal quantum gates and  can be written explicitly as follows
\[
H=\frac{1}{\sqrt{2}}
\left[ \begin{matrix} 1 & \ \ \  1 \cr 1 & -1\end{matrix}\right], \quad
T= \left[ \begin{matrix}1 &  0 \cr 0 & \e^{i\pi/4}\end{matrix}\right], \quad
\mathrm{CNOT}=
\left[ \begin{matrix}1 & 0 & 0 & 0 \cr 0 & 1 & 0 & 0 \cr  0 & 0 & 0 &  1 \cr  0 & 0 & 1 & 0 \end{matrix}\right] ,
\]
where $H$ and $T$ are given in the basis $\{|0\ra, |1\ra \}$ while
CNOT is in the two-qubit computational basis $\{|00\ra, |01\ra,  |10\ra, |11\ra\}$.

Unfortunately the decoherence and quantum noise due to local interactions destroying coherent phenomena and flipping $|0\ra \leftrightarrow |1\ra$ create unavoidable obstacles to building 
an efficient scalable quantum computation platform.
There are hopes that the quantum error-correcting algorithms could help creating 
efficient quantum computers from qubits and quantum gates that are not perfect,
however, these might not be a big help on the scale of 1000-qubit systems.
That is why it has become increasingly important to look for systems which are intrinsically immune against noise and decoherence. The topological quantum computers are good candidates 
for almost noiseless processing of quantum information.
%%%%%%%%%%%%%%%%%%%%%%%%%%%%%%%%%%%%%%%%%%%%
\section{What is Topological Quantum Computation?}
%%%%%%%%%%%%%%%%%%%%%%%%%%%%%%%%%%%%%%%%%%%%
Because the quantum noise and decoherence are presumably due to local interactions
we could try to avoid them by encoding quantum information non-locally,
for example, by using some topological characteristics of the system
such as, homotopy classes of quasiparticles exchange paths.
In this case quantum information is inaccessible to local interactions, because they
cannot distinguish between $|0\ra$ and $|1\ra$ and hence cannot
lead to decoherence and noise, providing in this way, a 
\textit{topological protection of qubit operations}.
Then the quantum gates should be some topologically non-trivial operations,
such as changing the homotopy classes, which could be implemented by braid 
operations (exchanges of quasiparticles in the plane).

However, in order to realize even a single topologically protected 
qubit, which belongs to a 2-dimensional space, we need state degeneracy  
in the plane and therefore higher dimensional 
representations of the braid group. The corresponding quasiparticles 
(known in the mathematical literature as plektons) are called non-Abelian anyons
and possess a number of very strange and interesting properties.

Perhaps the most promising non-Abelian candidate is the FQH state with filling 
factor $\nu=5/2$ that is routinely observed  in ultrahigh-mobility samples 
\cite{xia,eisen} in the second Landau level,
and is believed to be in the universality class of the Moore--Read Pfaffian state
whose two-dimensional  conformal field theory $\widehat{u(1)}\times \mathrm{Ising}$ 
contains the critical Ising model with central charge $1/2$.

The topological protection in this case comes from the fact that the residual noise and decoherence are due to thermally activated quasiparticle--quasihole (that might execute uncontrolled braidings), which are exponentially suppressed at low temperature
($T\sim 5$ mK) by the bulk energy gap ($\Delta\sim 500$ mK)\cite{xia,eisen}.
In Ref.~\cite{sarma-freedman-nayak} the error rate has been estimated to be
\[
 \mathrm{Error \ rate} \simeq  \left(\frac{k_B T}{\Delta}\right)
\exp\left(-\frac{\Delta}{k_B T}\right) \sim 10^{-30}
\]
which would lead to unprecedented precision of quantum information processing.

%%%%%%%%%%%%%%%%%%%%%%%%%%%%%%%%%%%%%%%%%%%%%%%%%%%%%%%%%%%%%%%%%%%%%%%%%%%%%
\section{Braiding: Extend TQC scheme of Das Sarma et al.}
%%%%%%%%%%%%%%%%%%%%%%%%%%%%%%%%%%%%%%%%%%%%%%%%%%%%%%%%%%%%%%%%%%%%%%%%%%%%%
The main idea of the TQC scheme of Ref.~\cite{sarma-freedman-nayak} is to use
as a qubit the state described by the Pfaffian wave functions with 4 quasiholes.
The quasiholes with coordinates  $\eta_1,\ldots, \eta_4$ are localized on 4 
antidots inside of an incompressible electron liquid formed by a large number 
 $N$ of  electrons (or holes) with coordinates $z_1, \ldots, z_N$ and the 
corresponding  wave function could be expressed as a chiral CFT correlation 
function
\beq\label{Psi}
\Psi_{4\qh}(\eta_1,\eta_2,\eta_3,\eta_4; z_i)
\mathop{=}\limits^{\mathrm{def}} \la \psi_{\mathrm{qh}}(\eta_1)\psi_{\mathrm{qh}}(\eta_2)\psi_{\mathrm{qh}}(\eta_3)
\psi_{\mathrm{qh}}(\eta_4)\prod_{i=1}^N
\psi_{\mathrm{h}}(z_i) \ra   ,
\eeq
of the CFT operators representing the holes and quasiholes 
\[
\psi_{\mathrm{h}}(z)= \psi(z) \, \np{\e^{i \sqrt{2}\phi(z)}  } \quad
\mathrm{and}\quad
\psi_{\mathrm{qh}}(\eta)= \s(\eta)\, \np{\e^{i\frac{1}{2\sqrt{2}} \phi(\eta)}},
\]
where $\psi(z)$ and $\s(\eta)$ are respectively the Majorana fermion and the 
chiral spin field of the Ising model and $\phi(z)$ is a normalized $u(1)$ boson.
The 4-quasihole wave function (\ref{Psi}) have been computed explicitly in Ref.~\cite{nayak-wilczek} to be
\[
\Psi_{4\mathrm{qh}}(\eta_1,\eta_2,\eta_3,\eta_4; z_1, \ldots, z_N)
=\Psi_{4\mathrm{qh}}^{(0)}+\Psi_{4\mathrm{qh}}^{(1)},
\]
where the functions $\Psi_{4\mathrm{qh}}^{(0,1)} \quad \leftrightarrow  \quad |0\ra ,| 1\ra $,
which will serve as a computational basis for the Pfaffian qubit, have the form
($\eta_{ab}=\eta_a-\eta_b$)
\[
\Psi_{4\mathrm{qh}}^{(0,1)}(\eta_1,\eta_2,\eta_3,\eta_4; z_1, \ldots, z_N)=
\frac{\left(\eta_{13}\eta_{24}\right)^{\frac{1}{4}}}{\sqrt{1\pm \sqrt{x}}}
\left(\Psi_{(13)(24)}\pm \sqrt{x}\,  \Psi_{(14)(23)} \right)
\]
and are expressed in terms of the single-valued Pfaffian functions 
\cite{nayak-wilczek,TQC-NPB}
\beqa
\Psi_{(ab)(cd)}&=&\Pf\left(
\frac{(z_i-\eta_a)(z_i-\eta_b)(z_j-\eta_c)(z_j-\eta_d)+
(i\leftrightarrow j)}{z_i-z_j}\right) \times \nn
&\times& \prod_{1\leq i<j\leq N} (z_i - z_j)^2 , \quad (a<b, \   c<d)   \nonumber
\eeqa
and $x=\eta_{14}\eta_{23}/ (\eta_{13}\eta_{24} )$ is the anharmonic CFT 
invariant crossratio.

The fermion parity is a subtle issue in the Ising model \cite{5-2,TQC-NPB}.
The Ramond superselection sector, is double degenerate
as a result of the presence of the Majorana fermion zero mode, i.e., there are two
chiral primary fields $\s_\pm$ with CFT dimension $1/16$, whose subscript is the 
chiral fermion parity $\gamma_F$
\[
[\psi_0,\gamma_F]_+=0, \quad \psi_0^2=\frac{1}{2}, \quad \gamma_F^2=1 \quad \Rightarrow \quad
\gamma_F \s_\pm \gamma_F= \pm\s_\pm .
\]
The physical quasihole, however, is represented by 
$
\s(\eta)=(\s_+(\eta)+\s_-(\eta))/\sqrt{2} 
$
because of the modular invariance requirement and the necessity of the so called
 GSO projection\cite{5-2}, which is actually the origin of its 
non-Abelian statistics expressed in the fusion rule $\s \times \s = \I + \psi$.

The conservation of the fermion parity implies that e.g., the correlation functions
involving 4 $\s$ fields and (possibly) an even number of Majorana fermions are non-zero
$\la \s_{e_1}\s_{e_2}\s_{e_3}\s_{e_4}\ra\neq 0$ only if  $e_1e_2=e_3e_4$.
On the other hand it follows from the Ising-model operator product expansion
that $\s_+\s_+ \simeq \s_-\s_-$ and $\s_+\s_- \simeq \s_-\s_+$, which together with $e_1e_2=e_3e_4$ reduces the dimension of the $4\s$-correlation functions space to  dim $\H_{4\s}=2$.
This allows us to write the computational basis in the CFT description as
\[
|0\ra \equiv \la \s_+\s_+\s_+\s_+\ra , \quad |1\ra \equiv \la \s_+\s_-\s_+\s_-\ra ,
\]
i.e., the state of the qubit is determined e.g., by the second pair $\s_{e_3}\s_{e_4}$ of
 $\s$ fields and the states of other pair is fixed by fermion parity conservation.
In general\cite{nayak-wilczek} the correlation functions space dimension for $2n$ $\s$ fields
 is  dim $\H_{2n\s}=2^{n-1}$, which means that $2n$ Ising anyons could be used to represent
$n-1$ qubits.
%%%%%%%%%%%%%%%%%%%%%%%%%%%%%%%%%%%%%%%%%%%%%%%%%
\section{Braid matrices from the Pfaffian wave functions with 4 quasiholes}
%%%%%%%%%%%%%%%%%%%%%%%%%%%%%%%%%%%%%%%%%%%%%%%%%
Now that we know how to write the computational basis in terms of Ising model 
correlation functions we could represent quasiparticle braidings, that would be used
as quantum gates, as analytic continuation of their coordinates. 
For example the counterclockwise 
braiding of the quasiparticles with coordinates $\eta_1$  and $\eta_2$ could be executed by the analytic continuation along the circle by
\[
\eta'_1= \frac{\eta_1+\eta_2}{2} + \e^{i\pi t } \frac{\eta_1-\eta_2}{2},  \quad
\eta'_2= \frac{\eta_1+\eta_2}{2} - \e^{i\pi t } \frac{\eta_1-\eta_2}{2} , \quad
0 \leq t \leq 1.
\]
Doing carefully the analytic continuation of the Ising-model 4-point functions
 $\sqrt{1\pm \sqrt{x}}$, with $x=\eta_{14}\eta_{23}/(\eta_{13}\eta_{24})$, and 
taking into account the contribution of the single-valued functions $\Psi_{(ab,cd)}$ 
we could obtain\cite{nayak-wilczek,TQC-NPB} the generators of the (positive parity) 
2-dimensional representation of  the braid group $\B_4$ to be
\beq \label{R4}
R_{12}^{(4)}=R_{34}^{(4)}=\left[ \begin{matrix}1 & 0 \cr 0 & i\end{matrix}\right],
\quad
R_{23}^{(4)}=
\frac{\e^{i\frac{\pi}{4}} }{\sqrt{2}}
\left[ \begin{matrix} \ \    1 & -i \cr -i & \ \ 1\end{matrix}\right] .
\eeq
Alternatively, these generators have been obtained in Ref.~\cite{slingerland-bais}
by using only quantum-group methods, which is more compact and perhaps more elegant. 
However, the direct derivation\cite{TQC-NPB} from the multi-quasihole Pfaffian wave
functions is necessary for many reasons.

\begin{figure}[htb]
\centering
\includegraphics*[bb=0 350 595 500,width=8cm]{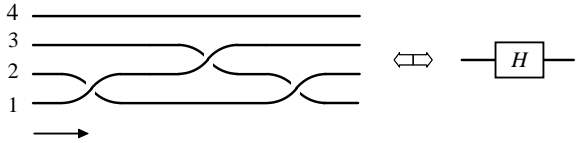}
\caption{Hadamard gate implemented by braiding 4 Ising anyons. \label{fig:H} }
\end{figure}

It turns out that the representation generated by Eq.~(\ref{R4}) is of finite order
 \cite{TQC-NPB}.
Given the explicit form of the generators of a finite group, we could 
enumerate all group elements  by using an optimized Dimino algorithm\cite{dimino}.
For the representation of the group $\B_4$ generated by (\ref{R4}) the
Dimino's algorithm gives (using Maple 10 with modules)
\[
|\mathrm{Image}\left(\B_4\right)| = 96.
\]
For general $2n$ Ising anyons the representation of the braid group $\B_{2n}$ 
(with $2n \geq 6$) is finite as well  and its order has been given in Ref.~\cite{Read-JMP}
 \[
|\mathrm{Image}\left(\B_{2n}\right)| = \left\{
\begin{array}{lll}2^{2n-1} (2n) ! & \quad \mathrm{for} & n=\mathrm{even} \\
2^{2n} (2n) ! & \quad \mathrm{for} & n=\mathrm{odd} \end{array} \right. .
\]
This is too bad because for universal TQC  we need 
$\mathrm{Image}\left(\B_{2n}\right)$ to be dense in the unitary group.
This means that not all quantum gates could be realized by braiding, hence, 
not all of them would be fully topologically protected.
Nevertheless, most of the important quantum gates, especially those 
from the Clifford group, can be implemented by braiding \cite{TQC-NPB}.
For example the single-qubit Hadamard gate can be constructed\cite{TQC-NPB} 
by 3 elementary exchanges as shown in Fig.~\ref{fig:H}.
\[
H\simeq R_{12}^2 R_{13} = R_{12} R_{23} R_{12}  =
\frac{\e^{i\frac{\pi}{4}} }{\sqrt{2}}
\left[ \begin{matrix}1 & \ \ \  1 \cr 1 & -1\end{matrix}\right].
\]
%%%%%%%%%%%%%%%%%%%%%%%%%%%%%%%%%%%%%%%%%%%%%%%%%%%%%%%%%%%%%%%%%%%%%%%%%%%%%%%
\section{Exchange matrices for $6$ Ising quasiholes: the two-qubit gates}
\label{sec:2qubits}
%%%%%%%%%%%%%%%%%%%%%%%%%%%%%%%%%%%%%%%%%%%%%%%%%%%%%%%%%%%%%%%%%%%%%%%%%%%%%%%
The two-qubit states could be represented by Pfaffian wave functions with 6 quasiholes 
localized on antidots \cite{TQC-NPB} and the computational basis would be e.g., 
\beqa
&&|00\ra \equiv \la \s_+ \s_+\s_+\s_+\s_+\s_+\ra, \quad
\! |01\ra \equiv \la \s_+ \s_+\s_+\s_-\s_+\s_-\ra \nn
&&|10\ra \equiv \la \s_+ \s_-\s_+\s_-\s_+\s_+\ra, \quad
|11\ra \equiv \la \s_+ \s_-\s_+\s_+\s_+\s_-\ra  . \nonumber
\eeqa

\begin{figure}[htb]
\centering
\includegraphics*[bb=10 340 580 490,width=\textwidth]{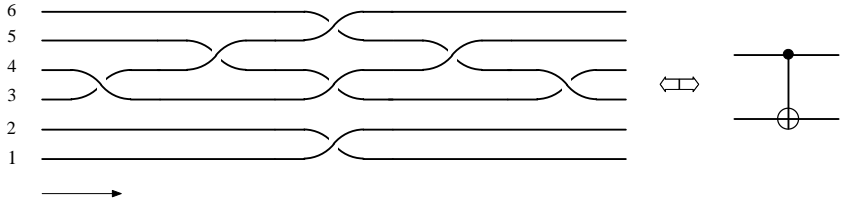}
\caption{Braid diagram for realization of the CNOT gate by braiding 6 Ising anyons.
The symbol to the right is the standard notation for CNOT. \label{fig:CNOT}}
\end{figure}

In order to obtain the braid matrices for exchanging 6 Ising anyons, which 
would represent the two-qubit gates, we don't
need to consider the full 6-quasiholes Pfaffian wave functions which are not 
even known explicitly. Instead,
we could use the general expression similar to Eq.~(\ref{Psi}), however with 6 quasihole 
operators $\psi_{\mathrm{qh}}$ inside the correlator. Then we can use the fact that 
the braiding operations are independent of the distance between the particles, especially for those which are  not  involved in the exchange, and we could fuse them to obtain a 4 quasihole Pfaffian wave function for 
which we could apply the results in Eq.~(\ref{R4}). For example, if we want to compute  $R^{(6)}_{12}$ then we could first fuse $\eta_5 \to \eta_6$ in which case the two-qubit computational basis could be written as
\beqa
&&|00\ra \mathop{\to}_{\eta_5 \to \eta_6}  \la \s_+ \s_+\s_+\s_+\ra, \quad
|01\ra \mathop{\to}_{\eta_5 \to \eta_6}   \la \s_+ \s_+\s_+\s_- \psi\ra \nn
&&|10\ra \mathop{\to}_{\eta_5 \to \eta_6}  \la \s_+ \s_-\s_+\s_-\ra, \quad
|11\ra \mathop{\to}_{\eta_5 \to \eta_6}  \la \s_+ \s_-\s_+\s_+\psi \ra  . \nonumber
\eeqa
Then the exchange $\eta_1 \leftrightarrow \eta_2$ is represented  by  $R_{12}^{(4)}$ 
and it is easy to see that
\[
R_{12}^{(6)} =  \mathrm{diag}(1,1,i,i) =R^{(4)}_{12} \otimes \I_2 .
\]
Similarly, by first fusing $\eta_5 \to \eta_6$, we could compute  $R^{(6)}_{23}$ as follows
\[
R_{23}^{(6)} = \frac{\e^{i\frac{\pi}{4}} }{\sqrt{2}} \left[
\begin{matrix}1 & 0 & -i & 0 \cr  0 & 1 & 0 & -i \cr -i & 0 & 1 & 0 \cr 0 & -i & 0 & 1\end{matrix} \right]
=R^{(4)}_{23} \otimes \I_2 .
\]
In the same way, however, by fusing first $\eta_1 \to \eta_2$, we  could obtain 
\[
R^{(6)}_{45}  =\frac{\e^{i\frac{\pi}{4}} }{\sqrt{2}}\left[
\begin{matrix} 1 & -i  & 0 & 0 \cr  -i & 1 & 0 & 0 \cr 0 & 0 & 1 & -i \cr 0 & 0 & -i & 1\end{matrix} \right]=
\I_2 \otimes R^{(4)}_{23} , \quad \mathrm{and}
\]
\[
R^{(6)}_{56}  = \mathrm{diag}(1,i,1,i)= \I_2 \otimes R^{(4)}_{34}.
\]
Unfortunately, this approach doesn't work for $R^{(6)}_{34}$ because, as we will see 
later, it is not factorizable due to the fact that the states of the 
quasiholes at $\eta_3$ and $\eta_4$ depend on both qubit's states.
However, due to a general argument \cite{TQC-NPB} based on the superselection rule
in the  Neveu--Schwarz sector, $R^{(6)}_{34}$ must be diagonal \cite{TQC-NPB}.
Therefore, we could obtain  $R^{(6)}_{34}$ directly from the  OPE for $\eta_3 \to \eta_4$
\beqa \label{fuse_34}
&&|00\ra \mathop{\to}_{\eta_3 \to \eta_4}  \eta_{34}^{-1/8}\la \s_+ \s_+\s_+\s_+\ra, \quad
|01\ra \mathop{\to}_{\eta_3 \to \eta_4}   \eta_{34}^{3/8}\la \s_+ \s_+ \psi \s_+\s_- \ra \nn
&&|10\ra \mathop{\to}_{\eta_3 \to \eta_4}  \eta_{34}^{3/8}\la \s_+ \s_- \psi \s_+\s_+\ra, \quad
|11\ra \mathop{\to}_{\eta_3 \to \eta_4}  \eta_{34}^{-1/8}\la \s_+ \s_-\s_+\s_- \ra  \nonumber
\eeqa
Taking into account the extra factor $\eta_{34}^{1/8} $ coming from the Abelian part of the quasiholes, and executing the exchange by $\eta_{34}\to \e^{i \pi} \eta_{34}$ we obtain 
the exchange matrix for $\eta_3 \leftrightarrow \eta_4$ to be
\beq \label{R6_34}
R^{(6)}_{34}  = \mathrm{diag}(1,i,i,1).
\eeq
It is worth stressing that the 6-quasihole exchange matrix $R^{(6)}_{34}$
cannot be expressed as a factorized tensor products of the 4-quasiholes exchange matrices and 
the unit matrix $\I_2$. This important fact, which is responsible for creating entanglement purely by braiding,  will be explained later in the universal $R$ matrix approach.
It is this topological entanglement which allows us to implement entangling two-qubit gates, such as the CNOT gate, purely by braiding 6 Ising anyons. Indeed, using the explicit formulae for the 6-quasiholes braid matrices, it is easy to check that the CNOT can be represented by 7 elementary exchanges as follows\cite{TQC-PRL,TQC-NPB} (see Fig.~\ref{fig:CNOT} for the corresponding braid diagram)
\[
\mathrm{CNOT}=R_{34}^{-1} R_{45} R_{34}R_{12}R_{56} R_{45}  R_{34}^{-1} .
\]
%%%%%%%%%%%%%%%%%%
%%%%%%%%%%%%%%%%%%%

%%%%%%%%%%%%%%%%%%%%%%%%%%%%%%%%%%%%%%%%%%%%%%%%
\section{The universal $R$ matrix approach}
%%%%%%%%%%%%%%%%%%%%%%%%%%%%%%%%%%%%%%%%%%%%%%%%%
The quantum group structure for the Pfaffian model has been analyzed in 
Ref.~\cite{slingerland-bais}.
Using the affine coset description, Ising$\equiv \widehat{su(2)}_2/\widehat{u(1)}$,
one could identify its quantum group as $U_q(sl(2))$ with $q=\e^{-i\frac{\pi}{4}}$.

As we have already mentioned the $n$-qubit states are represented by $2(n+1)$ 
Ising anyons which realize\cite{nayak-wilczek} spinor irreducible representations 
of $SO(2n+2)$ of dimension $2^n$.
Therefore the spinor representation of the braid-group generators could be expressed in 
terms of the spinor representation  of the universal $R$-matrix for $U_q(sl(2))$ with $q=\e^{-i\frac{\pi}{4}}$ as follows
\beq \label{tensor}
R_{k,k+1}^{(2n+2)} = \underbrace{\I_2 \otimes \cdots \otimes \I_2}_{k-1} \otimes R
\otimes \I_2 \otimes \cdots \otimes \I_2
\eeq
The spinor representation of the $R$ matrix for the Pfaffian FQH state 
(or for the Ising model) depends on the parity of $k$ 
\cite{nayak-wilczek,slingerland-bais,ivanov}. When $k$ is odd the $R$ matrix acts 
only on the $k$-th factor and has the form
\[
R = \left[
\begin{matrix}1 & 0 \cr 0 & i\end{matrix} \right],
\]
while for even $k$ the  $R$ matrix acts on the $k$-th and $(k+1)$-th factor as
\[
R = \frac{1}{\sqrt{2}}
\left[
\begin{matrix}1 & 0 & 0 &  -i \cr 0 & 1 & -i  &  0 \cr 0 & -i & 1 & 0
	\cr -i & 0 & 0 &  1 \end{matrix}\right] .
\]
Notice however, that the exchange matrix (\ref{tensor}) has dimensions 
$2^{2n+2} \times 2^{2n+2}$ while the spinor representations of $SO(2n+2)$ 
have dimension $2^n$. In order to obtain the action of the (irreducible) 
exchange matrices in the spinor representation with positive parity we have to apply the corresponding projector.
For example, in the case of 6 Ising anyons, using the obvious  notation $\s_+\s_+ \sim +$
and $\s_+\s_- \sim -$, 
we have to project the states 
\[
|+++\ra, |++-\ra, |+-+\ra, |+--\ra, |-++\ra, |-+-\ra, |--+\ra, |---\ra
\]
to the two-qubit basis $\{ |+++\ra, |+--\ra, |-+-\ra, |--+\ra \} $ with positive parity.
Therefore, to obtain the braid-group generators in the representation  we have to
apply the projector
\[
P_2:=\mathrm{diag}(1,0,0,1,0,1,1,0).
\]
Now let us go back to the braid matrix for the exchange of $\eta_3$ and $\eta_4$ in the 6-anyon representation. Applying the projector $P_2$  and deleting all null rows and columns 
we get the same result as Eq.~(\ref{R6_34})
\[
R^{(6)}_{34}=P_2 \left( \I_2\otimes \left[\begin{matrix}1 & 0 \cr 0 & i\end{matrix}\right]\otimes \I_2 \right) P_2 =\mathrm{diag}(1,i,i,1) .
\]
Obviously factorizability of the braid matrix (inside the parentheses above) is 
lost after projection and this projection is the origin of the topological 
entanglement discussed in Sect.~\ref{sec:2qubits}.

It is worth mentioning that the Artin's relation for the spinor representations of the
braid-group generators follow from the Yang--Baxter equation for the $R$ matrix
only if the projectors commutes with
$\I_2 \otimes \cdots \otimes \I_2 \otimes R \otimes \I_2 \otimes \cdots \otimes \I_2$,
however, it turns out that for the Ising anyons this is always the case.

Finally it might be interesting to note that this representation of the braid 
matrices in terms of the $R$ matrix is a nice illustration of the 
idea of Kauffman et al. to use universal $R$ matrices for TQC \cite{kauffman-braid}.
Of course we have to be aware that some specific details, such as parity projectors 
creating topological entanglement and the impossibility to realize by braiding 
all single-qubit gates, may be model-dependent, however, we have to admit that the 
opportunity for topological quantum entanglement due to braiding of non-Abelian anyons 
is remarkable.

%%%%%%%%%%%%%%%%%%%%%%%%%%%%%%%%%%%%%%%%%%%%%%
\section{Conclusions}
%%%%%%%%%%%%%%%%%%%%%%%%%%%%%%%%%%%%%%%%%%%%%%
Several open problems should be mentioned: first of all it might be very useful
if we could construct the missing $\pi/8$ gate by braiding plus something else 
in such a way to obtain maximal protection from noise. 
Alternatively one may try to construct a protected
 Toffoli gate. The embedding of all single-qubit and two-qubit gates into
three-qubit system still has to be resolved. Another question is how to 
approximate the  quantum Fourier transform more efficiently and finally it would be 
necessary to analyze the possible error sources in this TQC scheme.

%%%%%%%%%%%%%%%%%%%%%%%%%%%%%%%%%%%%%%%%%%%%%%%%%%%%%%%%%%%%%
% Doing Acknowledgements                                     %
%%%%%%%%%%%%%%%%%%%%%%%%%%%%%%%%%%%%%%%%%%%%%%%%%%%%%%%%%%%%%

\section*{Acknowledgments}
I would like to thank Ivan Todorov, Ady Stern,  Valentina Petkova, Chetan Nayak,
Lyudmil Hadjiivanov, Michael Geller and Preslav Konstantinov for many helpful 
discussions.
The author has been supported as a Research Fellow by the Alexander von Humboldt
foundation. This work has been partially supported by the BG-NCSR under Contract
No. F-1406.
%%%%%%%%%%%%%%%%%%%%%%%%%%%%%%%%%%%%%%%%%%%%%%%%%%%%%%%%%%%%%
% Doing references:                                         %
%%%%%%%%%%%%%%%%%%%%%%%%%%%%%%%%%%%%%%%%%%%%%%%%%%%%%%%%%%%%%
%\bibliography{Z_k,my,TQC,FQHE}
%\begin{thebibliography}{10}
\providecommand{\href}[2]{#2}\begingroup\raggedright\endgroup

\end{document}